\documentclass[aps,prl,twocolumn,groupedaddress,superscriptaddress]{revtex4}

\usepackage{graphicx}  
\usepackage{dcolumn}   
\usepackage{bm}        
\usepackage{amssymb}   
\usepackage{amsmath}
\usepackage{xypic,color}
\usepackage[para]{threeparttable}
\begin{document}


\title{On the dephasing of genetic oscillations.}


\author{D. A. Potoyan}
\affiliation{Center for Theoretical Biological Physics, Rice University, Houston, TX 77005, USA}
\affiliation{Department of Chemistry, Rice University, Houston, TX 77005, USA}
\author{P. G. Wolynes}
\email{pwolynes@rice.edu}
\affiliation{Center for Theoretical Biological Physics, Rice University, Houston, TX 77005, USA}
\affiliation{Department of Chemistry, Rice University, Houston, TX 77005, USA}
\affiliation{Department of Physics \& Astronomy, Rice University, Houston, TX 77005, USA}



\date{\today}

\begin{abstract}

The digital nature of genes combined with the associated low copy numbers of proteins regulating them is a significant source of stochasticity, which affects the phase of biochemical oscillations. We provide a theoretical framework for understanding the dephasing evolution of genetic oscillations by combining the  phenomenological stochastic limit cycle dynamics and the discrete Markov state models that describe the genetic oscillations. Through simulations  of the realistic model of the $NF\kappa B$/$I\kappa B$ network we illustrate the dephasing phenomena which are important for reconciling  single cell and population based experiments on this system.

\end{abstract}

\pacs{}

\maketitle

\section{Introduction}

\begin{figure}[!ht]
\centering
  \includegraphics[scale=0.23]{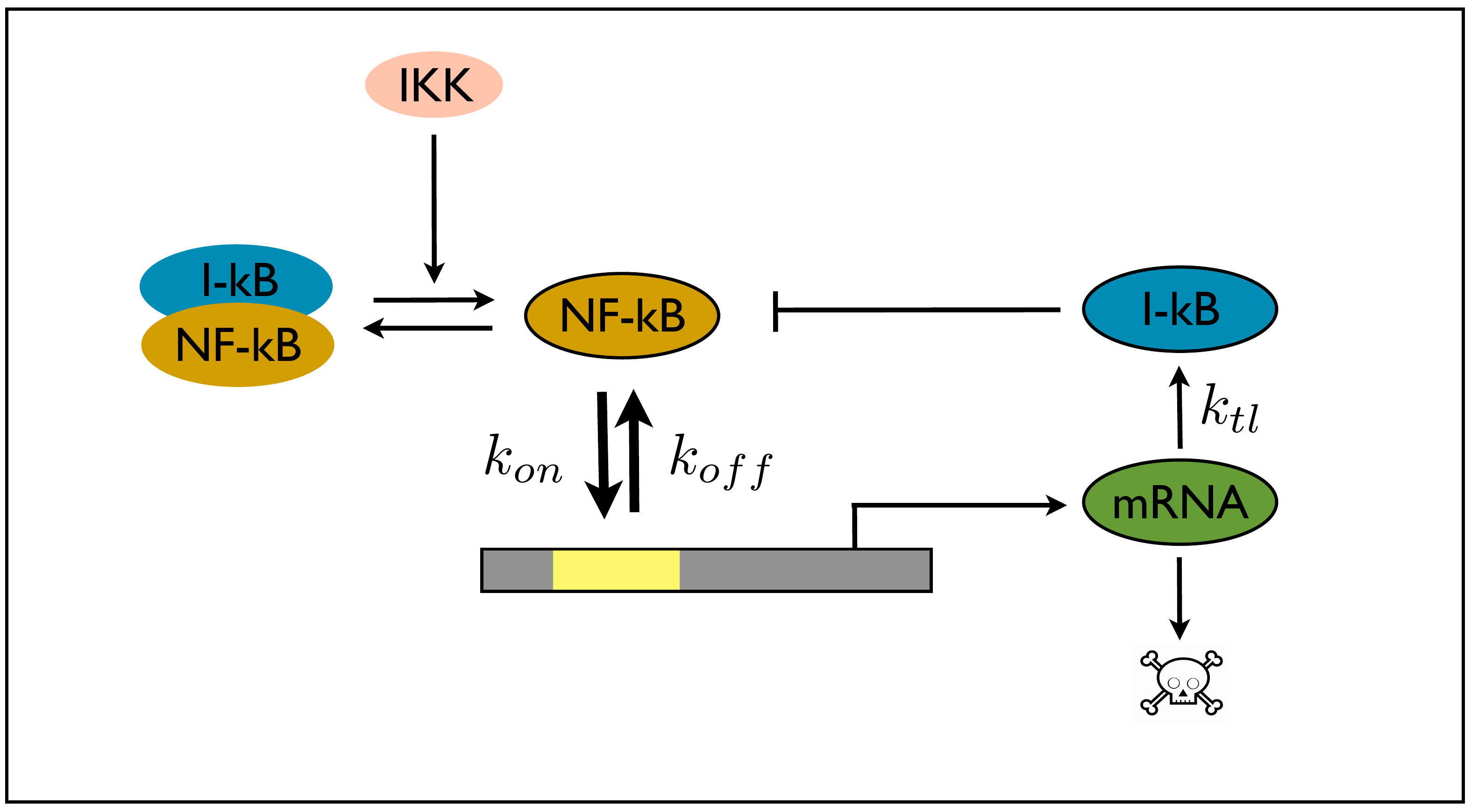}
  \caption{A minimalist model of $NF\kappa B$/$I\kappa B$ genetic oscillator. Bold arrows indicate binding ($k_{on}$) and unbinding ($k_{off}$) of $NF\kappa B$ to the gene. Once bound, $mRNA$ is produced which initiates the synthesis ($k_{tl}$) of the $I\kappa B$. The copies of $mRNA$ are also constantly degraded. The $I\kappa B$ inhibits the $NF\kappa B$ by binding to it and preventing the activation of the gene. The $IKK$ drives the irreversible degradation of $I\kappa B$ in the complex and prevents the full de-activation of $NF\kappa B$.}
  \label{fig:intro}
\end{figure}

Cyclic dynamics is a common feature of many self-organized systems~\cite{winfree2001geometry} manifesting itself in myriad  forms in biology, ranging from sub-cellular biochemical oscillations to cell division and on to the familiar predator-prey cycles of ecology. An oscillatory response of a gene regulatory circuit, whether transient or self-sustained can have a number of advantages over a temporally monotonic response~\cite{schultz2009deciding,morelli2007precision}. The ability of copies of a system to synchronize can lead to dramatic noise reduction and greater precision to the timing in assemblies. On the sub-cellular level rhythmic dynamics spans time scales from a few seconds as in the calcium oscillations to days as in the circadian rhythms or years for cicada cycles~\cite{goldbeter1997biochemical, winfree2001geometry, novak2008design}. The ultradian genetic oscillations, which take place at an intermediate scale from minutes to a few hours are medically important. A singularly important example is the $NF\kappa B$ gene network, which organizes a cell's response to various types of external stress and plays a role in regulating inflammation levels in populations of cells. The response of the $NF\kappa B$ circuit to continuous external stimulation has been studied by Hoffmann et al~\cite{hoffmann2002ikappa} who observed damped oscillatory dynamics for $NF\kappa B$, which has been linked to the presence or absence of particular forms of the inhibitor $I\kappa B$. On the other hand, experiments carried out on individual cells have detected more sustained $NF\kappa B$/$I\kappa B$ oscillations which are either completely self-sustained~\cite{ashall2009pulsatile,nelson2004oscillations} or damp at a much slower rate~\cite{nelson2004oscillations} than found for the population depending on the duration of external stimulation. It follows that some type of averaging takes place, but the physical mechanisms and the stochastic aspects of this population averaging are not fully understood. Many aspects of the $NF\kappa B$ oscillatory dynamics can be rationalized using deterministic mass action rate equations~\cite{krishna2006minimal,tiana2007oscillations}, but how stochastic self sustained oscillations average out at a cell population level still remains unanswered. In this work we provide a conceptual framework for understanding of stochastic averaging as a result of ``dephasing" of genetic oscillators.  We explore a particular simple yet realistic model of the $NF\kappa B$/$I\kappa B$ circuit (Fig.~\ref{fig:intro}) and demonstrate how self-sustained stochastic single cell oscillatory dynamics yields damped oscillations at the population level as observed in the experiments of Hoffmann et al~\cite{hoffmann2002ikappa}.  Another related but more fundamental question is how the single molecule nature of the gene contributes to the stochasticity of the network. In our model, we explicitly account for the highly non-Gaussian noise coming from a single gene turning on/off and investigate how the timescale of the DNA operator state fluctuations impacts the noisiness of the oscillatory dynamics. The ideas and techniques developed in the present work should be broadly applicable for studying dephasing effects in other genetic oscillators. 

{\it Phenomenological model of dephasing.\textemdash } By dephasing, one essentially means the loss of common phase or coherence of oscillations in populations of $mRNA$ or protein byproducts of gene activation, caused by the stochastic events in the course of an oscillator's operation that occur at different times in different cells~\cite{yoda2007roles}. In the near deterministic limit, self-sustained oscillations can be described with an autonomous system of equations with limit cycle attractors. For such an attractor trajectories  rapidly relax towards the limit cycle but once on it the phase undergoes ``free diffusion" driven by the underlying fluctuations. Let us first discuss the deterministic model of self-sustained oscillator which is born via a supercritical Andronov-Hopf (AH) bifurcation~\cite{guckenheimer1983nonlinear}. Classical examples~\cite{goldbeter1997biochemical,novak2008design,winfree2001geometry} of chemical oscillations created via AH bifurcation include the Belouzov-Zhabotinsky reaction, Brusellator, Selkov model of glycolysis and many others. The normal form of the Andronov-Hopf bifurcation may be written~\cite{guckenheimer1983nonlinear}:
\begin{equation}
\dot z = p_1(\Lambda - \Lambda_c)  z-p_3|z|^2z,
\end{equation}
Where z is the representation of the limit cycle in the complex plane in the vicinity of the bifurcation. In the context of gene circuits, z is the dynamical variable which describes the oscillatory cycle formed by any two components in the feedback loop. Such a simple description is a consequence of planarity of the attractor in the phase space. The parameters $p_1$ and $p_3$ are complex expansion coefficients $p_n=p_n^{'}+ip_n^{''}$, $\Lambda$ is the bifurcation control parameter and $\Lambda_c$ is its critical value. Transforming the variables to polar form, one obtains the normal forms for amplitude $\dot r = (\Lambda-\Lambda_c) p^{'}_1 r - p^{'}_3 r^3$, and phase $\dot \phi = (\Lambda-\Lambda_c) p^{''}_1  - p^{''}_3 r^2$. Due to the attractor nature of the limit cycle the amplitude equilibrates on a faster time scale, while the phase evolves freely. Therefore we may put  $r$ at its equilibrium value $r_0=\left(\frac{(\Lambda-\Lambda_c)p^{'}_1}{p^{'}_3} \right)^{1/2}$, so that the long time behavior of phase will be: 
\begin{equation}
\phi -\phi_0 = (\Lambda-\Lambda_c) \left[ p^{''}_1  -  \frac{p^{'}_1}{p^{'}_3}p^{''}_3\right]t = \omega t. 
\end{equation}
Deterministic limit cycle dynamics can be described as an oscillation with a constant angular frequency $\dot \phi = \omega_0$, and amplitude $r=r_0$. 
The accuracy of the deterministic picture deteriorates as we scale down the size of our system so that at some point the fluctuating molecular nature of our oscillator must be accounted for~\cite{yuan2013exploring,baras1982asymptotic}.  
As a first order approximation we can describe the dynamics of phase and amplitude as being driven by a Gaussian white noise $\eta(t)$,
\begin{equation}
\dot \phi(t) = \omega_0 + v +\sqrt{2D_{\phi}} \cdot \eta(t),
\end{equation}
\begin{equation}
\dot r(t) = \alpha r -\lambda r^{3} +\sqrt{2D_{r}} \cdot \eta(t). 
\end{equation}
Where $v= \langle \dot \phi \rangle - \omega_0$ is the noise induced shift of the phase, $D_r$ and $D_{\phi}$ are the diffusion coefficients for amplitude and phase and the $\lambda$ and $\alpha$ are constant parameters which depend on the rate coefficients in the gene oscillator network. It is convenient to work with the corresponding Fokker-Plank equation $\partial_t P(r, \phi, t)  = \hat L_{FP} P(r, \phi, t)$. Denoting  $\omega=\omega_0+v$, the Fokker-Planck operator reads:
\begin{equation}
 \hat L_{FP}  = \frac{\partial }{\partial r} \left(-\alpha r +\lambda r^{3} +D_{r}\frac{\partial }{\partial r} \right)+\frac{\partial }{\partial \phi} \left(\omega +D_{\phi}\frac{\partial }{\partial \phi} \right).
\end{equation}
Since the dynamics of the amplitude is decoupled from phase diffusion the $P(r,\phi,t)$ becomes simply a product of two individual distributions
$P(r,\phi,t)=p^{ss}(r) p(\phi,t)$. Since we also assume that amplitude dynamics takes place at a faster time scale we may use the steady state probability distribution satisfying the equation, $J^{ss}_r = \frac{\partial }{\partial r} \left(-\alpha r +\lambda r^{3} +D_{r}\frac{\partial }{\partial r} \right)p^{ss}(r) = 0$. The phase distribution is seen to be that of a simple Wiener process and is given by standard form of Green solution to the diffusion equation with $\phi(t)\in\mathbb{R}$ modulo $2\pi$. Taking the time evolution from a specified initial condition of phase $ P(r,\phi,t) = e^{L_{FP}t} \delta(\phi-\phi_0) $, one obtains the following form for the probability distribution:
\begin{equation}
 P(r,\phi,t) = \frac{1}{N(t)} e^{\frac{1}{D_r} \left (\frac{\alpha}{2} r^2- \frac{\lambda}{4} r^4 \right)} e^{-\frac{(\phi-\omega t-\phi_0)^2}{4 D_{\phi}t}}, 
\end{equation}
where the $N^{-1}(t)$ is a normalization factor. At this point it is a straightforward to compute the correlation function by averaging time lagged product of observables, $C(\tau)=\langle r_{\tau} cos\phi_{\tau} \cdot r_0 cos\phi_0\rangle$. As phase diffusion is decoupled from amplitude fluctuations we can perform the averages separately and take the product at the end. The radial part yields $\langle r \rangle^2 = \int \int dr dr_0 p^{ss}(r) p^{ss}(r_0) r  r_0 dr = \frac{\pi D_r}{16 \lambda} e^{\alpha^2/2 \lambda D_r}\left[1+erf\left(\alpha \sqrt{\frac{1}{4 \lambda D_r}}\right)\right]^2$. For reasonable values of parameters $\langle r \rangle^2 \sim D_r$, implying that noise increases the effective length of the oscillation amplitude. For the phase it is more instructive to derive the correlation function by first writing the phase, $\phi(t)- \omega t -\phi(0) =\int_{0}^{t} \eta(t) dt$.  In the complex exponential $\langle e^{i(\phi(t)-\phi(0))} \rangle$ representation, after  taking the average over realizations of noise $\eta(t)$ using the gaussian gaussian white noise assumption one finds $\langle cos\phi_t \cdot cos\phi_0 \rangle = \frac{1}{2} e^{-D_{\phi} t} cos(\omega t)$.
Combining the last two expressions for amplitude and phase  we obtain the final expression for the correlation function: 
\begin{equation}
C(\tau) =  \frac{1}{2} \langle r \rangle^2 e^{-D_{\phi} \tau } cos\omega \tau.
\label{eq:phencorr}
\end{equation}
The correlation is a damped cosine oscillating at an average stochastic frequency $\omega$ with a constant average amplitude. In stochastic dephasing, the damping time scale is set by the noise intensity $D_{\phi}$ of the phase variable. It represents a ``virtual" damping, since it results from dephasing of the trajectories due to stochastic fluctuations while individual trajectories themselves would appear to continue to oscillate. 
   
{\it Discrete state Markov models of dephasing.\textemdash} The modeling of the phase and amplitude evolution via Langevin dynamics introduces noise into the system in an {\it ad hoc}  manner appropriate to a near macroscopic system. For genetic oscillators one can take a different path by starting from a microscopic, discrete state description provided by the master-equation. The state of the system $S$ is given by specifying the number of all proteins together with the occupation of states of the genes. Once transition rates $W_{ij}$ between the pairs of states $i,j \in {\it S}$ are assigned the probability ket $ |P(t) \rangle$ can be found starting from a given initial condition $|P(t=0) \rangle$. Assuming Markovian dynamics for the transitions we have:
\begin{equation}
\partial_t |P(t)\rangle = {\bf W} |P(t)\rangle\\.
\end{equation}
The elements of the stochastic rate matrix $\bf W$, are the transition rates between states $W_{ij} =  \langle i | {\bf W}| j \rangle$ for $i \neq j$ and the net escape rates from specified states $i$,  $W_{ii} =  \langle i | {\bf W}| i \rangle = - \sum_i \langle i | {\bf W}| j \rangle$.
The probability of a state, $z$ is given by $p(z,t|z_0,0) = \langle z|P(t)\rangle $. 
Owing to time translation invariance  we can write the solution using the eigenvectors $(|V^{(i)}\rangle)$ and eigenvalues $(\lambda_i)$ of the rate matrix ${\bf W}$. Assuming the eigenvalues are not degenerate the general solution is $|P(t)\rangle = e^{{\bf W}t} |P(0)\rangle =  \sum^{N-1}_{i=0} e^{{\lambda_i}t} |V^{(i)}\rangle \langle V^{(i)} |P(0)\rangle$. 
The Perron-Frobenius theorem requires the existence of at least one purely real eigenvalue with real part zero, while the rest of the eigenvalues must have strictly negative real parts ($\lambda_0=0, Re(\lambda_i)<0, i=1,...N-1 $). For the systems with broken detailed balance, the rate matrix is non-hermitian~\cite{qian2000pumped}, which may produce complex eigenvalues corresponding to oscillatory motion along $N_c$ cycles for which rations like $\frac{W_{12} W_{23}...W_{N_c 1}}{W_{21} W_{32}...W_{1N_c}} \neq 1$ hold~\cite{qian2000pumped,schnakenberg1976network}. Circular fluxes do not average out when system size is scaled, which leads to limit cycles on macroscopic scales~\cite{wang2008potential}. The conditional probabilities for states $z$ are obtained by taking the scalar product of $|P(t)\rangle$ with bras: $p(z,t|z_0,0)=\sum^{N-1}_{i=0} \langle z | V^{(i)} \rangle e^{\lambda_i t} \langle V^{(i)} | P(0) \rangle $. The stationary state is obtained by simply taking the limit of very long time: $ p^{ss}(z)= p(z,\infty |z_0,0)=  \langle z | V^{(0)} \rangle  \langle V^{(0)} | P(0) \rangle = \langle z | V^{(0)} \rangle $.
The last equality follows from the fact that the left eigenvector corresponding to stationary state is unity, $\langle V^{(0)} | ={\bf 1}$. 
The correlation function is given by $C(\tau) = \langle z_\tau z_0 \rangle = \int dz_{\tau} \int dz_0 p(z_{\tau}| z_0) p^{ss}(z_0) z_\tau z_0$. Plugging in the expressions for the probabilities and integrating over variables $z_{\tau}$ and $z_0$ yields:
\begin{equation}
C(\tau) =  \alpha^2_0 + \sum^{N}_{i=1} e^{\lambda_i t} \alpha_i \beta_i.
\end{equation}
Where the $\alpha$'s and $\beta$'s are the integrals, $\alpha_i = \int  \langle z_{\tau} | V^{(i)} \rangle z_{\tau} d z_{\tau}$, and $\beta_i = \int  \langle z_{0} | V^{(0)} \rangle \langle V^{(i)} | P(0) \rangle z_0 d z_{0}$. Since ${\bf W}$ has real elements, the eigenvalues and eigenvectors come in complex conjugate pairs, i.e. $\alpha_i \beta_i=(\alpha_{i+1}\beta_{i+1})^{*}$ and $\lambda_i = \lambda^{*}_{i+1}$. The pair with a smallest $\gamma$ corresponds to the most slowly evolving mode(s). Assuming a significant spectral gap between first and the rest of the exponents and subtracting the constant stationary fluctuation term, $\alpha^2_0 = \langle z \rangle_{ss}^2$, one obtains: 
\begin{equation}
C(\tau)  \approx R_0  e^{-|\gamma| \tau}  cos(\omega \tau+\phi_0).
\label{eq:mastercor}   
\end{equation}
Where the $R_0= |\alpha \beta|$ and $\phi_0=arg(\alpha \beta)$ are constants set by the initial condition. The slowest modes, according to our assumption are linked with the motion along the cycle, while the fast ones are due to the fluctuations longitudinal to the cycle.
\begin{figure}[!ht]
\centering
  \includegraphics[scale=0.22]{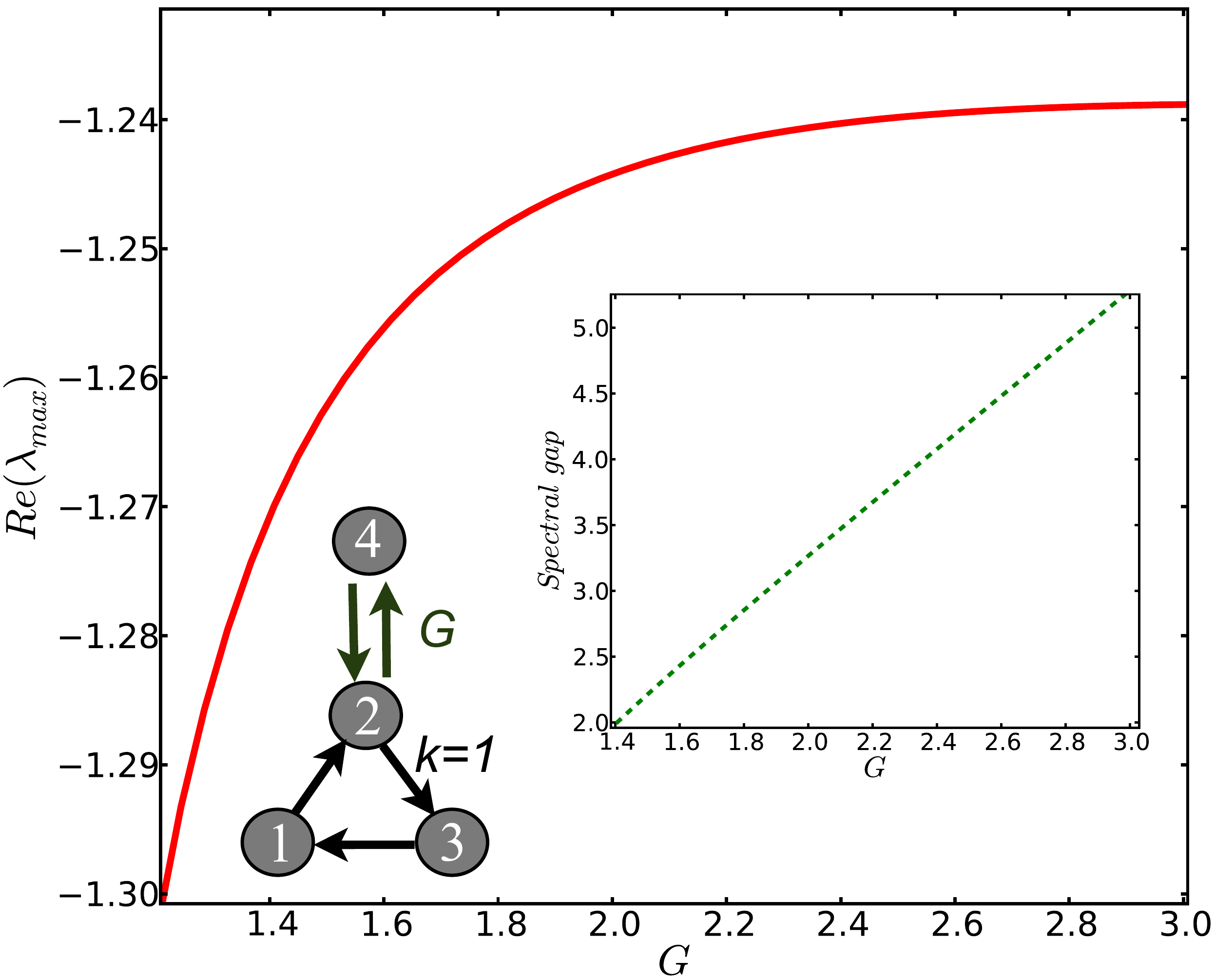}
  \caption{A mock up model of genetic oscillator: a three state directed
cycle with an attached edge. The graph shows the dependence of the real part of the largest non-trivial eigenvalue on the switching rate (G), with the cycling rates (k) fixed. The inset shows the dependence of spectral gap on G.}
   \label{fig:eig}
\end{figure}
The apparent similarity of the form of the correlation function (\ref{eq:mastercor}) to that of (\ref{eq:phencorr}) provides a justification for employing the framework of the phenomenological approach, which was based on rather restrictive approximations. In contrast, in the master equation formalism one only invokes the much milder assumption that there exists a sufficient spectral gap in the eigenvalues. The last assumption can be readily tested on mock up models of genetic oscillators. To that end we propose a model of a directed cycle with a unit attached edge (See Fig \ref{fig:eig}). The role of the edge is in mimicking the binary nature of gene states, which moderate the cyclic dynamics to an extent, which depends on the switching timescale. By increasing the switching rate, the real part of largest non-trivial eigenvalue increases eventually saturating at a high switching rates $G \gg k$, implying that dephasing dynamics becomes $G$ independent. Qualitatively similar behavior is seen with our more realistic stochastic simulations which is elaborated in the next section.   

 
 {\it Stochastic simulations of $NF\kappa B$/$I\kappa B$ oscillator model.\textemdash} At the simplified level, the core of the NF$\kappa$B gene network consists of a negative feedback (Fig~\ref{fig:intro}) by the inhibitors $I\kappa B$, which suppress the binding of the NF$\kappa$B to the gene that activates the synthesis of the I$\kappa$B. The cyclic suppression of its own synthesis by the $I\kappa B$ creates a closed time delayed loop, resulting in oscillations of the components of the network.  Our model of the network is based on the scheme proposed by Sneppen et al~\cite{krishna2006minimal,tiana2007oscillations}, who explored the oscillatory behavior of $NF\kappa B/I\kappa B$ using a system of deterministic ODEs. A key difference between their system and ours is in the explicit incorporation of the digital nature of gene activation (OFF+NF$\kappa$B $\rightleftarrows$ON, see Fig.1) as opposed to modeling transcription as a first order mass action kinetics. We stochastically simulate the system of reactions via a kinetic Monte Carlo algorithm~\cite{gillespie1977exact} (see SI for details). 
\begin{figure}[!ht]
\centering
  \includegraphics[scale=0.25]{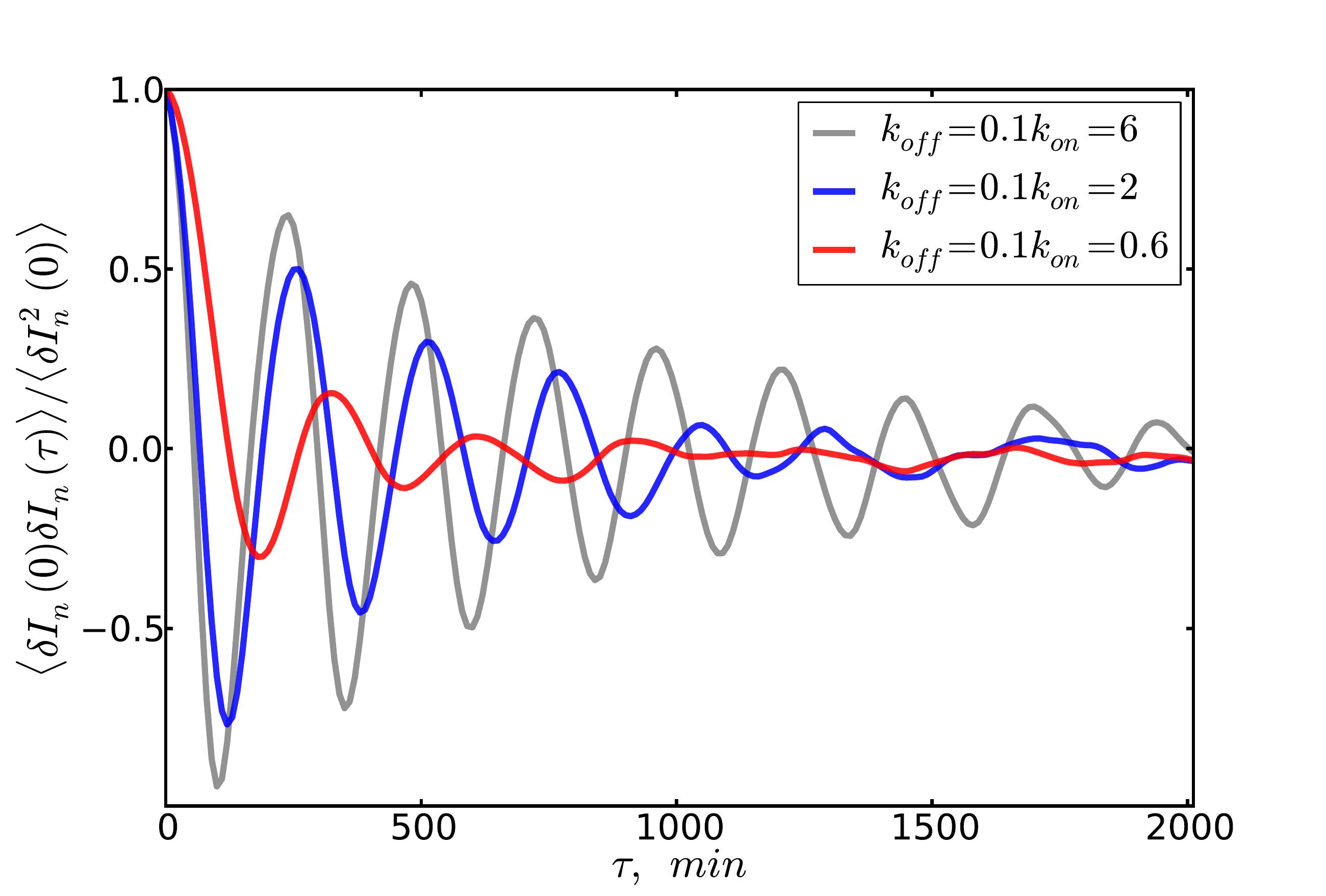}
  \caption{Normalized autocorrelation of steady state fluctuations in $I\kappa B$ population as a function of binding/unbinding rate.}
  \label{fig:one}
\end{figure}
The single molecule nature of the gene in our model turns out to be quite crucial as we find that the time scale of operator binding and release has a significant impact on the dephasing times $\tau_{\phi}$, which can be seen by the change in the decay rate of the correlations  $e^{-\tau/\tau_{\phi}}cos(\omega \tau)$ of the $I\kappa B$ (Fig.~\ref{fig:one}). The slowing of operator binding/release transitions boosts the dichotomic gene noise, which increases the phase diffusion $\langle \Delta \phi^2(t)\rangle \sim D_{\phi} t$ (Fig.~\ref{fig:one}, SI), or lowers the real part of the eigenvalue corresponding to oscillatory mode (Fig.~\ref{fig:eig}), leading to the faster damping of the oscillations. One may achieve the same levels of dephasing as were observed in the experiments of Hoffmann et al~\cite{hoffmann2002ikappa} by simply slowing the binding ($k_{on}$) and unbinding ($k_{off}$) rates, while maintaining the ratio equal to its known~\cite{bergqvist2009kinetic} {\it in vitro} value $(K_d=k_{off}/k_{on}=0.1)$. This provides an alternative explanation for the damping of the oscillations from that found using the deterministic models which suggest the experimentally observed damping must arise through the action of the other members of the $I\kappa B$ family via some independent reaction pathways.   
\begin{figure}[!ht]
\centering
  \includegraphics[scale=0.22]{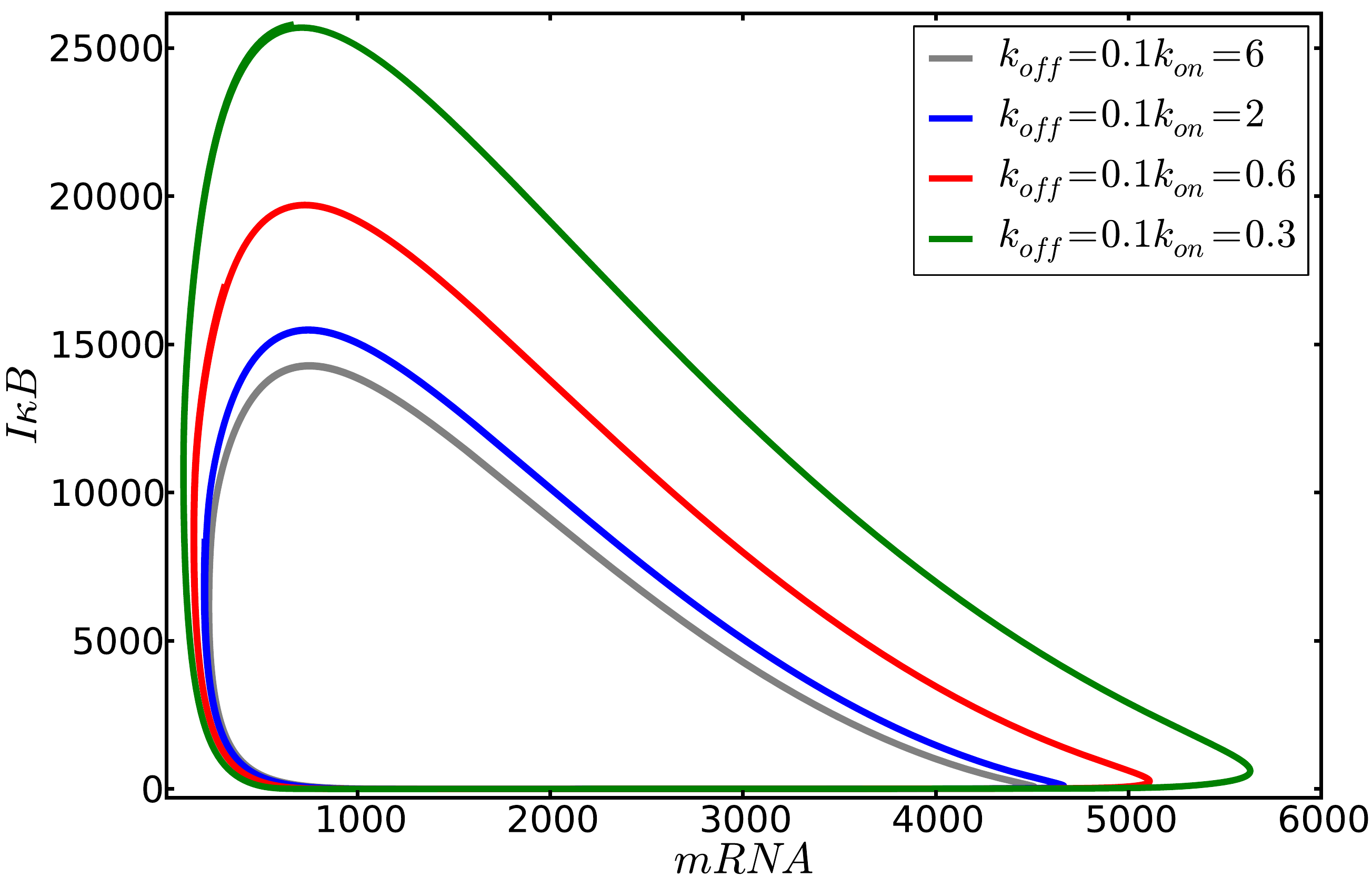}
  \caption{Radial averages of the stochastic I$\kappa$B/$mRNA$ limit cycles for different binding/unbinding rate coeficients. Axes are in the units of numbers of molecules.}
   \label{fig:lcycle}
\end{figure}
 Besides making the oscillations more stochastic, the slowing of gene state changes, also delays the feedback by the $I\kappa B$. The delay is reflected in the expansion of the limit cycle (Fig.~\ref{fig:lcycle}), or lowering of the oscillation frequencies, as is seen from the systematic leftward shift of the power spectrum peak (see SI). Interestingly the noise induced expansion of the limit cycles is also qualitatively captured by the phenomenological model, $\langle r \rangle^2 \sim D_r$ which predicts higher oscillation amplitudes with more ``noisy" oscillators. 
At last, one expects the rate of dephasing to be a function of oscillation cycling rate or the period. 
\begin{figure}[!ht]
\centering
  \includegraphics[scale=0.22]{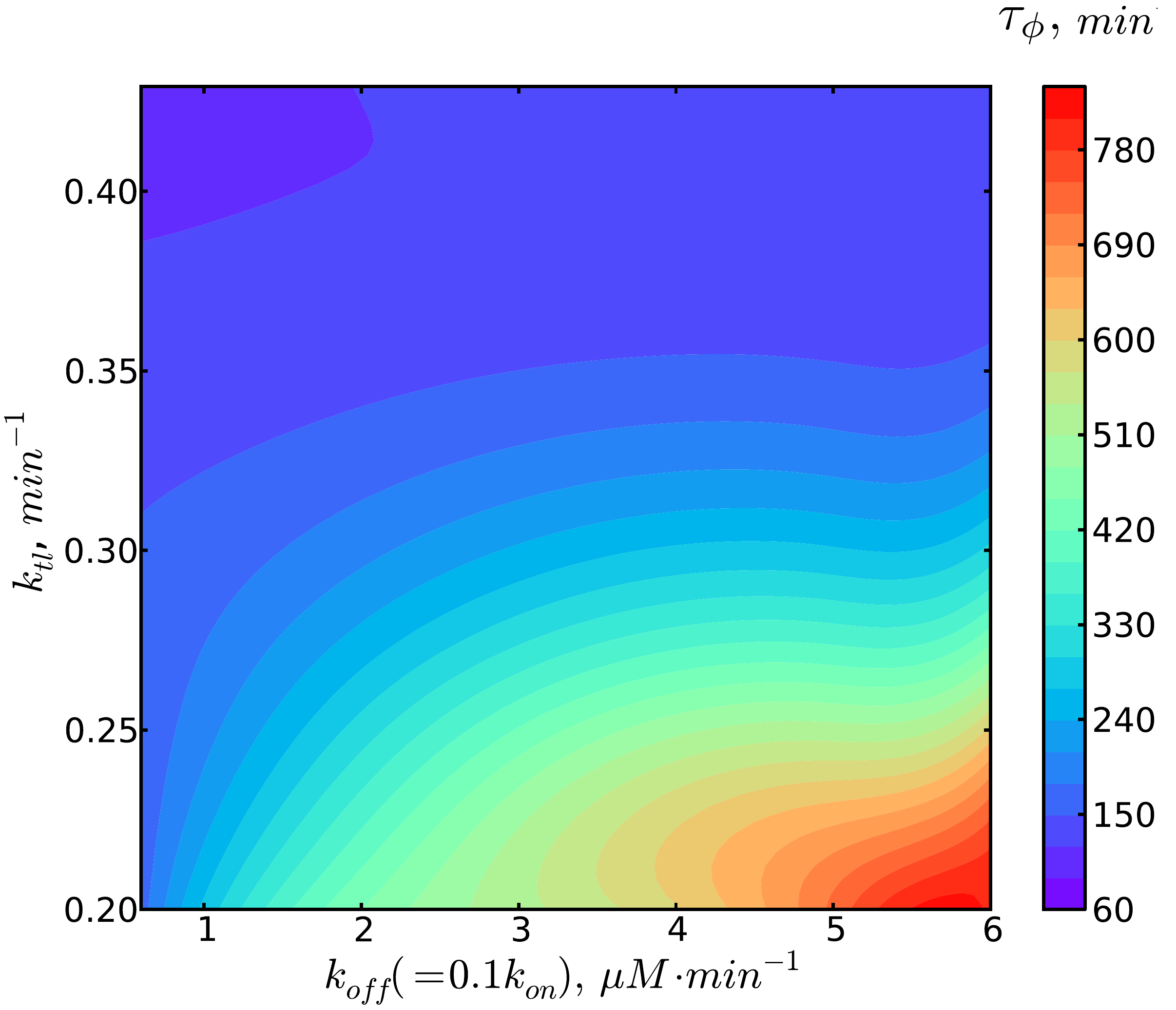}
  \caption{Dependence of the dephasing time $\tau_{\phi}$ on the rates of translation ($k_{tl}$) and gene state switching ($k_{off}).$}.
  \label{fig:2D}
\end{figure}  
 In our model we can tune the cycling rate by adjusting the irreversible rates  that the are part of the negative feedback loop of the network, while keeping the binding/unbinding rates fixed. By doing so one finds that both faster cycling rate and slower binding/unbinding rates do indeed lead to faster dephasing and vice versa (Fig~\ref{fig:2D})). Thus, the factors of noise and cycling rate are seen as the main contributors to the dephasing, where the former quantifies the rate of de-correlation but only the later is needed to set the time scale of the oscillations. 
\newline Financial support by the D.R. Bullard-Welch Chair at Rice University and PPG grant P01 GM071862 from the National Institute of General Medical Sciences are gratefully acknowledged.

\bibliography{ref.bib}
\section{Supplementary Materials}
\newpage
\newpage

The model of $NF\kappa B/I\kappa B$ oscillator consists of 14 elementary reaction events which are presented in the Table I. The stochastic simulations are done via kinetic Monte Carlo algorithm of Gillespie~\cite{gillespie1977exact}. All ensemble averages are computed by running the simulations $\sim 10^4$ times for the duration of $\sim$3500 min each and using the last 3000 min of trajectories, where the non equilibrium steady state has been established. All simulations were initiated from the same initial state with $\sim10^3$ molecules of $NF\kappa B$ and in the absence of all other proteins and $mRNA$. The simulation mimics the experiment of Hoffmann et al~\cite{hoffmann2002ikappa} where the cells are under constant exposure of external stimuli, which results in steady production of $IKK$ and keeps the single cell oscillations undamped.  

\begin{figure}[!ht]
\centering
  \includegraphics[scale=0.36]{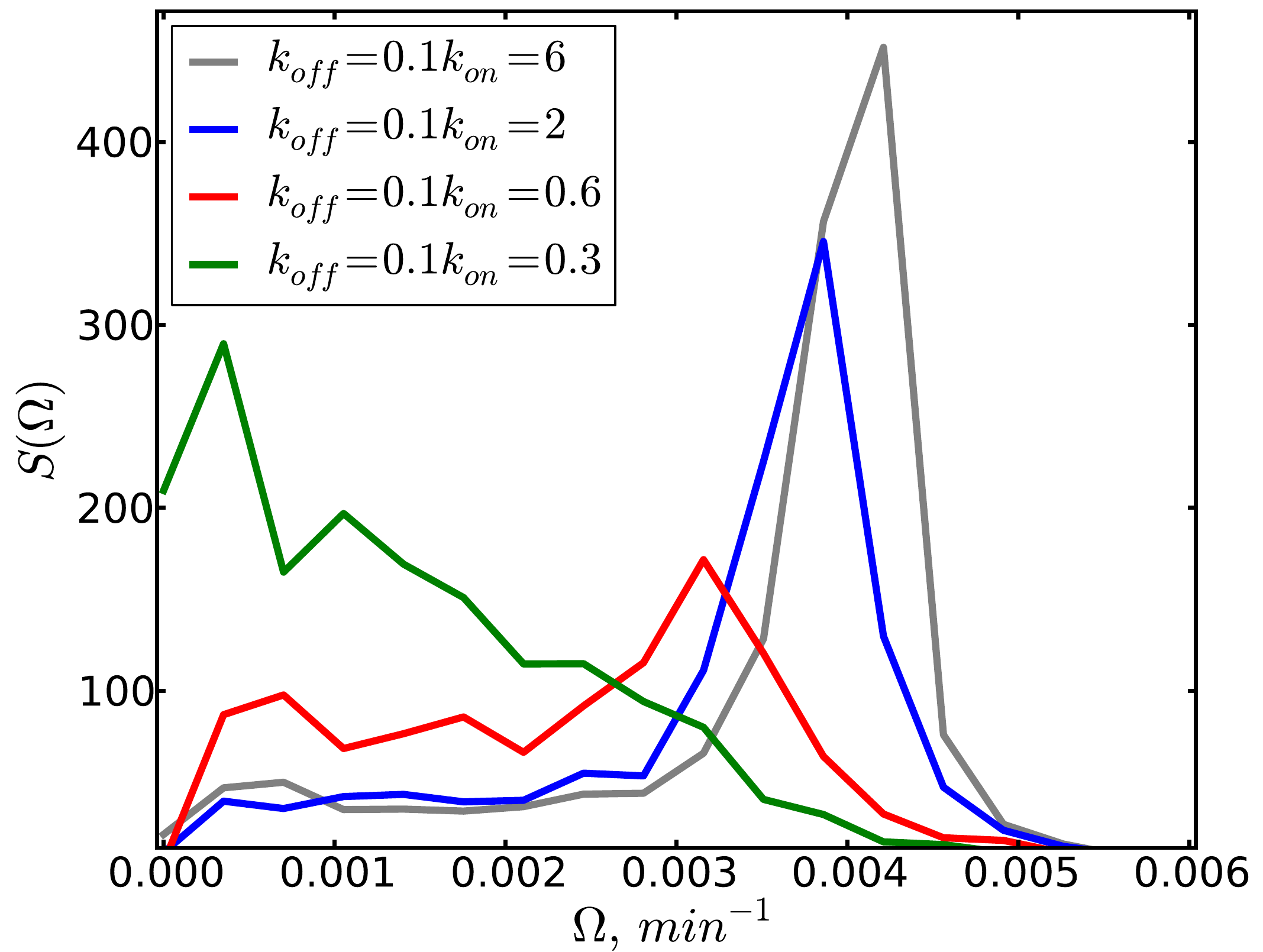}
  \caption{The power spectrum of the $I\kappa B$'s stochastic oscillations for different binding/unbinding rate coeficients. In the  phenomenological phase diffusion model it is given by $S(\Omega) =  \frac{D_{\phi}\langle r \rangle^2}{2} \left(  \frac{1}{D^2_{\phi}+(\Omega+\omega)^2} +\frac{1}{D^2_{\phi}+(\Omega-\omega)^2} \right)$. The slower operator state fluctuations  broaden the power distribution which corresponds to higher values of dephasing rate $D_{\phi}\sim\frac{1}{\tau_{\phi}}$ and therefore higher levels of noise in the system.}
   \label{fig:pow}
\end{figure}

\begin{table}[!ht]
\addtolength{\tabcolsep}{-8pt}
\begin{threeparttable}
 \caption{Reactions and rate coefficients for the NF-kB network. The ON and OFF indicate states of the gene. Labels (nuc) and (cyt) refer to nuclear and cytoplasmic concentrations. The $[N:I]$ standss for the complex between $NF\kappa B$ and $I\kappa B$. The irreversible reactions are indicated by $\Rightarrow$ arrow. }
\begin{tabular}{@{\vrule height 10.5pt depth4pt  width0pt}cccccc}
&\multicolumn5c{\hspace{40 mm} }\\
\noalign{\vskip-11pt}
\cline{1-6}
\vrule depth 0pt width 0pt Reactions &{ \hspace{0 mm} Rate Coeff  }&{ \hspace{0 mm}Values}\\
\hline{}
$OFF+NF\kappa B_{ (nuc)} {\rightarrow}  ON$ &$k_{on}  $&60.0 $min^{-1}$& \\
$ON \rightarrow OFF+NF\kappa B_{(nuc)}$  &$k_{off} $&6.0 $\mu M \cdot min^{-1}$&\\
$ON \Rightarrow ON+mRNA$  &$k_t $&1.03 $\mu M \cdot min^{-1}$&\\
$mRNA \Rightarrow mRNA + I\kappa B_{(cyt)}$  &$k_{tl}$&0.24 $min^{-1}$&\\
$mRNA \Rightarrow \emptyset$  &$\gamma_m$&0.017 $min^{-1}$&\\
$I\kappa B_{(cyt)} \rightarrow I\kappa B_{(nuc)}$ &$k_{Iin}$&0.018 $min^{-1}$&\\
$I\kappa B_{(nuc)} \rightarrow I\kappa B_{(cyt)}$ &$k_{Iout}$&0.012 $min^{-1}$&\\
$NF\kappa B_{(cyt)} \Rightarrow NF\kappa B_{(nuc)}$ &$k_{Nin}$&5.4 $min^{-1}$\\
$NF\kappa B_{(cyt)} +I\kappa B_{(cyt)} \rightarrow [N:I]_{(cyt)}$ &$k_f $&30.0 $\mu M^{-1} \cdot min^{-1}$&\\
$ [N:I]_{(cyt)} \rightarrow  NF\kappa B_{(cyt)} +I\kappa B_{(cyt)}$ &$k_b $&0.03 $min^{-1}$&\\
$NF\kappa B_{(nuc)} +I\kappa B_{(nuc)} \rightarrow [N:I]_{(nuc)}  $ &$k_{fn} $&30.0 $\mu M^{-1} \cdot min^{-1}$&\\
$ [N:I]_{(nuc)} \rightarrow NF\kappa B_{(nuc)} +I\kappa B_{(nuc)} $ &$k_{bn}$&0.03 $min^{-1}$&\\
$[N:I]_{(cyt)} \Rightarrow NF\kappa B_{(cyt)} $&$\alpha$& 0.55 $min^{-1}$&\\
$[N:I]_{(nuc)} \Rightarrow [N:I]_{(cyt)} $&$k_{out}$& 0.83 $min^{-1}$&\\
\hline
\end{tabular}
\end{threeparttable}
\end{table} 

\end{document}